\documentclass[showpacs,preprintnumbers,prl,twocolumn]{revtex4}

\usepackage{amssymb}
\usepackage{amsmath}
\usepackage{graphicx}
\usepackage{subfigure}
\usepackage{epsfig}

\begin{document}

\title{Measuring order in the isotropic packing of elastic rods}

\author{E. Bayart, S. Deboeuf, F. Corson, A. Boudaoud\dag and M. Adda-Bedia}

\affiliation{Laboratoire de Physique Statistique, Ecole Normale Sup\'erieure, UPMC  
Paris 06, Universit\'e Paris Diderot, CNRS, 24 rue Lhomond, 75005 Paris,  France.}

\date{\today}

\begin{abstract}

The packing of elastic bodies has emerged as a paradigm for the study of macroscopic disordered systems. However, progress is hampered by the lack of controlled experiments. Here we consider a model experiment for the isotropic two-dimensional confinement of a rod by a central force. We seek to measure how ordered is a folded configuration and we identify two key quantities. A geometrical characterization is given by the number of superposed layers in the configuration. Using temporal modulations of the confining force, we probe the mechanical properties of the configuration and we define and measure its effective compressibility. These two quantities may be used to build a statistical framework for packed elastic systems.

\end{abstract}


\maketitle

Packed elastic objects are ubiquitous in Nature and technology. For instance, DNA is packed in the cell nucleous or in viral capsids \cite{Arsuaga2002, Katzav2006}, while growing tissues can be confined by their environment \cite{DarcyBook, Couturier2009}. The optimization of folding is crucial in the design of self-deployable structures, such as tents or solar sails \cite{Miura1980}, or in waste disposal. Like a granular pile, a confined plate can be either in a crystalline state, the stacked facets obtained by repeatedly folding a sheet into two, or in a disordered state, exemplified by a crumpled ball \cite{Tallinen2008}. When a sheet is confined, the number of metastable configurations blows up \cite{Boue2006}. Meanwhile, self-avoidance leads to jamming because it prevents the system from exploring the space of configurations. This raises the question of whether a confined sheet can be viewed as a glassy system, in the same class as a static granular medium \cite{Jaeger1996}. Indeed, theoretical studies proposed thermodynamical approaches for packed rods \cite{Katzav2006, Boue2007, Aristoff2010}. On the experimental side, a difficulty in the study of crumpled balls \cite{Gomes1987, Matan2002, Blair2005, Andresen2007, Balankin2007, Lin2009} arises from the hand-generation of configurations. In this context, the confinement of a rod in a plane was an important and useful simplification \cite{Donato2002, Boue2006, Stoop2008, Deboeuf2009}. However, drawing general conclusions from these experiments can be questioned because of issues such as friction between the periphery of the configuration and the container, the anisotropic injection of the rod in the container, plasticity or the impossibility of unfolding a configuration. 

Here we reconsider the packing of a rod in a plane. We devised an experiment allowing us to reversibly confine a rod by a central force, deriving from an isotropic radial potential. As a consequence, there is no contact between the container and the periphery of the configuration, while the intensity of the forcing is controlled through the stiffness of the potential. We investigate the emergence of geometrical order through the stacking of layers. As this setup enables the temporal modulation of the confinement, we also probe the mechanical properties of configurations, and define an effective compressibility of a configuration, which we associate with geometrical order. We thus obtain a coupled geometrical and mechanical characterization of the system.

\begin{figure}[htbp]
\centering
\includegraphics[height=7.5cm]{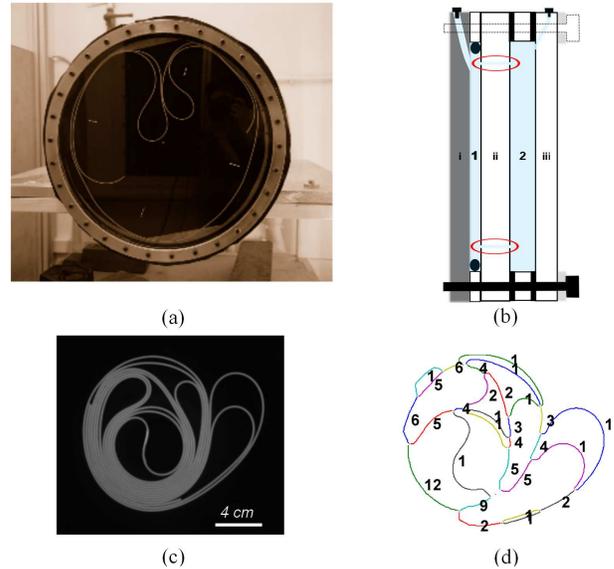}
\caption{The experiment. (a) A rod of density 1 is confined in a circular Hele-Shaw cell, filled with saturated salted water of density 1.16. The rod is in its initial configuration, before the cell has been set in rotation. (b) The cell is made of a superposition of three disks; the intermediate one ($ii$) is pierced with two holes to enable equilibration of pressure into the liquid. The chambers are made watertight with flat and toric joints shown in dark. (c) Example of a configuration obtained when a centripetal force is generated by the rotation of the cell around its axis. (d) Corresponding skeletonized image in which the number of layers per branch is defined.}
\label{exp}
\end{figure}

The principle of the experiment is as follows. A circular Hele-Shaw cell is filled with a liquid and entrained by a motor. A rod is inserted into the cell. The cell is slightly thicker than the rod, so that the rod cannot cross itself, constraining a two-dimensional folding. The liquid is denser than the rod. As a consequence, when the cell is rotated, the rod is submitted to a centripetal force and thus confined in a radial, parabolic potential. The cell was fixed in a vertical position on the axis of a 2kW motor, through a double ball bearing to avoid vibration transmission from the motor to the cell. The rotation velocity was controlled using an electronic frequency variator. We used a PDMS rod of circular section, $h=2\pm\,0.2\,$mm in diameter and $L=3\pm0.01\,$m in length and density $\rho=1$. The liquid was salt-saturated degassed water of density $\rho_l=1.16$ at ambient temperature. The cell was initially made of two disks of $50\,$cm in diameter: a stiff one in $10\,$mm--thick Duralumin ($i$), and a transparent one in $15\,$mm--thick Polycarbonate ($ii$), allowing us to observe the confined rod (Fig.~\ref{exp}a,b). Chamber 1 was filled with the liquid and the rod; the gap thickness was fixed at $2.5\,$mm by a ring of Plexiglas inserted between the two disks. However it turned out that this first design does not allow a constant gap. Indeed, the rotation induced a pressure gradient in the liquid, with a negative pressure at the center, leading to the bending of the Polycarbonate plate and the thinning of the gap. The solution was to add a second Polycarbonate disk ($iii$) to obtain a second chamber of thickness $1.5\,$cm; this chamber is filled with the same liquid and communicates with chamber 1 with two holes of $1\,$mm in diameter, enabling the equilibration of pressure in the two chambers. Chamber 2 serves as a sacrificial chamber: when the cell is rotated, disk ($iii$) bends inwards (instead of disk ($ii$)) but the gap is thick enough so that it does not close at our maximal rotation velocity of  $20\,$Hz. The chambers were filled through holes pierced in the edge of disks ($i$) and ($iii$) and the rod was gently inserted using a hole in the back of disk ($i$). The whole setup was placed in a dark room and lit with three stroboscopic lamps with diffusing screens. Movies were taken with a CCD camera. The duration of a flash was short enough ($1\,$ms) to get sharp images of the rod, even at the higher velocities. Using a computer interface, the camera and stroboscopes were triggered with the same square periodic signal, while the variator was controlled using a DC voltage. To enhance contrast, we used a white rod on a dark back: an adhesive sheet of black plastic was laid on the Duralumin disk ($i$). Binary images of the rod were obtained by thresholding.

\begin{figure}[htbp]
\centering
\includegraphics[height=9cm]{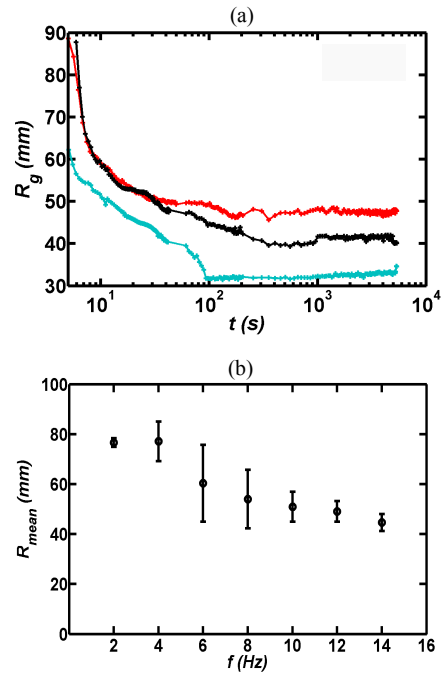}
\caption{The radius of gyration. (a) The radius of gyration $R_g$ as a function of time for three realizations in which the cell was launched from 0 to $14\,Hz$ in $3\,s$. Radii reach a plateau value in about $10^3\,$s. The final radius differs according to realizations. (b) On average, the radius of gyration decreases with the strength of confinement (quantified by the rotation velocity, $f$ of the disk). The error bars are sized according to the standard deviation of measured radii for each frequency.}
\label{rgt}
\end{figure}

The initial configuration of the rod (Fig.~\ref{exp}a) is prepared using 8 magnetic beads of diameter $1.8\,$mm inserted into the first chamber and moved from outside with a magnet. The cell is then set in rotation; the time to reach the desired frequency is approximately $3\,$s. The control parameter is the frequency $f$ of rotation (See EPAPS Document No. [number will be inserted by publisher] for a movie of a typical experiment.) First, configurations can be characterized using the radius of gyration, 
\[
R_g^2=\frac{1}{L}\int_0^Lr^2(s)\,\mathrm{d}s\mathrm{,}
\]
where $s$ is the curvilinear coordinate of the rod and $r(s)$ the distance to the cell axis. This quantity is directly calculated using the binary image of the confined rod. The radius of gyration decreases with time, rapidly when the rotation is started (Fig.~\ref{rgt}a) and then more slowly, reaching a plateau value in a time lapse of the order of $10^3\,$s. This final value is not unique and differs according to realizations at a given frequency. As we only investigated equilibrium configurations, we had to wait $1800\,$s for each value of the frequency before taking measurements, to ensure that equilibrium is reached. When averaging over realizations, the radius of gyration is found to be a decreasing function of confinement strength (Fig.~\ref{rgt}b). It appears that, when the same experiment is repeated, a large diversity of sizes and geometries is observed. The aim of the present study is to quantify geometrical order in equilibrium states. To do so, we extract the skeleton of a folded configuration from binary images. Vertices are detected as self-contact points, ie points of the skeleton having three neighbors (Fig.~\ref{exp}c, d). We define branches as portions delimited by two vertices. A given branch may contain several layers of the rod: the thickness of the branch on the binary image directly yields the number of layers. 

\begin{figure}[htbp]
\centering
\includegraphics[height=8cm]{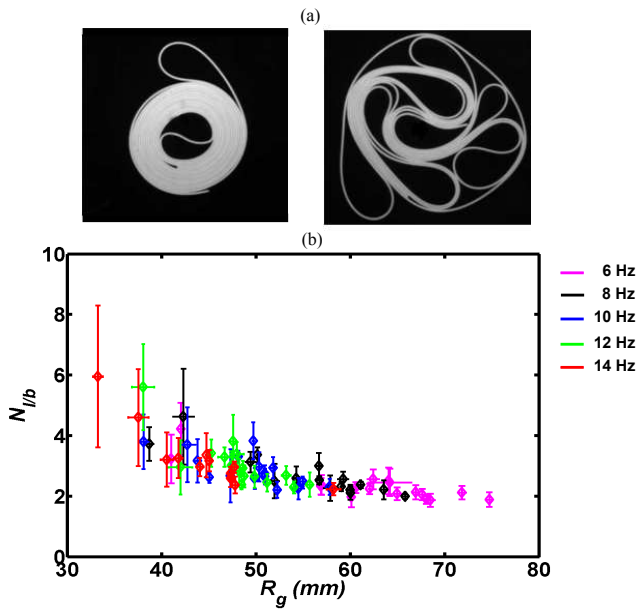}
\caption{Stacking. (a) Order and disorder. Two extreme examples of equilibrium configurations obtained at two different frequencies, with a large and a small number of superposed layers, respectively. (b) Average number of layers per branch, $N_{l/b}$, as a function of the radius of gyration $R_g$. Each point corresponds to one realization of the experiment. The colors of the symbols correspond to the imposed rotation velocity of the disk. An ordered configuration (large $N_{l/b}$) has a small radius of gyration.}
\label{layers}
\end{figure}

Inspired by observations (Fig.~\ref{layers}a), we first consider a geometrical definition of order using the number of superposed layers. Indeed, the configuration of absolute minimum of energy is a spiral~\cite{Boue2006}, which can be qualified as very ordered and in which all layers are superposed and contained in a unique branch. Experimentally, this number of superposed layers is defined as the average number of layers per branch in a given configuration. We jointly measured the average number of layers per branch $N_{l/b}$ and the radius of gyration of a configuration, as shown in Fig.~\ref{layers}b. The data roughly collapse on a single curve, independently of the confinement strength. Therefore, either the radius of gyration or the number of layers appear as better characterizations of the configuration. A disordered configuration, ie with small $N_{l/b}$, has a larger radius than a spiral ordered configuration. As in some previous experiments~\cite{Deboeuf2009, Lin2009}, stacking appears as a distinctive feature of the confinement of rods, and our setup allows us to show that stacking decreases with the radius of the configuration. In the following we investigate the possibility of characterizing geometrical order without a detailed knowledge of the geometry of configuration. In other words, we seek an independent measure of order.

\begin{figure}[htbp]
\centering
\includegraphics[height=10cm]{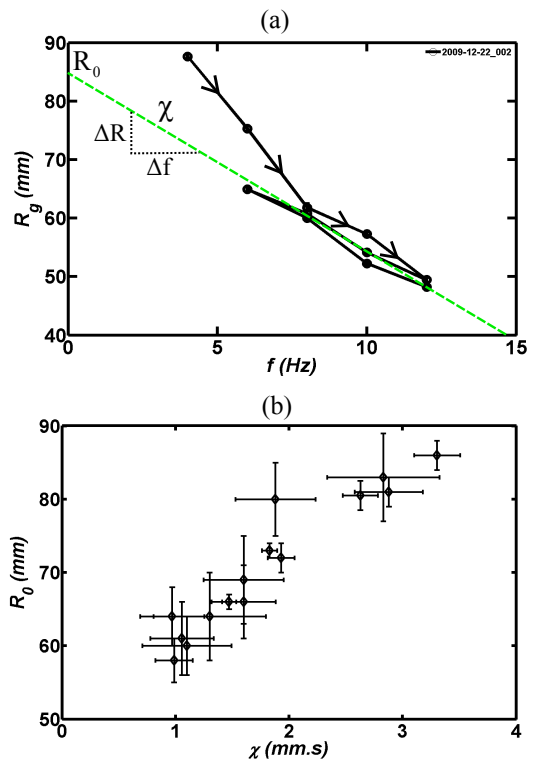}
\caption{Compressibility. (a) Evolution of the radius of gyration $R_g$ during an annealing experiment. After a first irreversible brach, the radius follows a reversible branch (parallel to the dashed line). A configuration is characterized by the slope of the line, $\chi$, and its intercept, $R_0$, with the axis $f=0$. (b) The characteristic radius $R_0$ as a function of compressibility $\chi$. Each point corresponds to one realization of an annealing experiment. The error bars correspond to the uncertainties estimates of the linear fits as described in (a).}
\label{annealing}
\end{figure}

\begin{figure}[htbp]
\centering
\includegraphics[height=2.2cm]{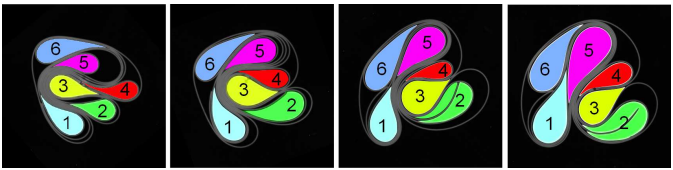}
\caption{Example of the evolution of a configuration during an annealing experiment. Rotation frequency was varied from $12\,$Hz (left) to $6\,$Hz (right), by steps of $2\,$Hz of $1800\,$s duration. Each colored area corresponds to a loop. The relative position of loops is invariant. The configuration only breathes under the perturbation.}
\label{topo}
\end{figure}

We realized annealing experiments in which the confinement strength was repeatedly increased then decreased by varying the rotation velocity appropriately. At each step in rotation frequency, we waited $1800\,$s in order to reach equilibrium (Fig.~\ref{rgt}a), so that an annealing experiment typically took 12 hours. After a few frequency steps, the radius of gyration followed approximately the same line (Fig.~\ref{annealing}a). The first steps are irreversible, while the line is reversible. In other words, the system follows an irreversible branch in the $(f,R_g)$ phase space before falling on a reversible branch. This behavior is reminiscent of the evolution of the density of a tapped granular pile according to the tapping acceleration, as reported in \cite{Nowak1997}. In our case, each annealing experiment can be characterized with the intercept, $R_0$, and with the slope, $\chi$, of the reversible branch. Furthermore, we observed that no topological changes occurred along the reversible branch, as illustrated in  Fig.~\ref{topo}: the relative positions of loops remain constant, and the configuration only seems to breath. As a consequence a given annealing experiment leads to a well-defined configuration, which can be characterized with $R_0$ and $\chi$. $R_0$ corresponds to its effective radius, while $\chi$ defines an effective compressibility of the configuration as it measures its susceptibility to the confinement strength. When plotting these two quantities, the characteristic radius $R_0$ is found to be an increasing, roughly linear function of the compressibility $\chi$ (Fig.~\ref{annealing}b). The results are independent of the confinement (frequency) since, whatever annealing path, points seem to collapse on the same master curve. As we found above that an ordered configuration has a small radius of gyration (Fig.~\ref{layers}b), this second set of results indicates that an ordered configuration has a small compressibility, and, conversely, a disordered configuration is highly compressible. A possible interpretation is that self-avoidance imposes a stringent constraint on ordered configurations for which more stacking means less freedom in exploring phase space; as a consequence an ordered configuration would be less compressible.

To summarize, we built an experiment allowing the two-dimensional confinement of a rod in a parabolic potential. This experiment allowed us to quantify order and disorder in a configuration, using a geometrical measure of stacking and a mechanical measure of (effective) compressibility. We found that these two quantities are strongly correlated with the characteristic radius of the configuration. Although this effective compressibility is strictly a susceptibility to changes in rotation velocity, it can be readily generalized to other systems such as crumpled balls. In that context, a salient feature of compressibility is that its measurement does not require a full knowledge of the geometry of the configuration. In other words, order could be inferred from the effective stiffness of the crumpled ball. Future work should address whether these quantities could be used in a thermodynamic approach to packing, or how these quantities could emerge in such an approach. Thus geometrical and mechanical properties appear as strongly entangled in the packing of sheets and rods.

We are grateful to J. Da Silva Quintas for his help in building the experimental apparatus.\\
\dag Current address: RDP, ENS Lyon, 46 all\'ee d'Italie, 69007 Lyon, France

\bibliography{compressibility}

\end{document}